\def\C(#1){|#1|}

\def\eqar#1{\begin{eqnarray} #1 \end{eqnarray}}
\def\eq#1{\begin{equation} #1 \end{equation}}
\def\eqn#1{\begin{equation} \nonumber #1 \end{equation}}

\def\mbit#1{\mbox{\boldmath$ #1 $}}
\def\Def{\triangleq}

\def\Ind(#1){\Delta\left[#1\right]}

\def\b{\beta}
\def\Inner(#1,#2){\langle #1, #2 \rangle}

\documentclass[journal, letterpaper, twocolumn, final]{IEEEtran}

\usepackage{amsmath}
\usepackage{amssymb}
\usepackage{graphicx}
\usepackage{graphicx}
\usepackage[noadjust]{cite}
\title{Iterative Algebraic Soft-Decision List Decoding of Reed-Solomon Codes
\thanks{This research was supported by
NSF grant no. CCR-0118670 and grants from Sony, Qualcomm, and the
Lee Center for Advanced Networking. The material in this paper was
presented in part at the International Symposium on Information
Theory and its Applications, Parma, Italy, October, 2004.}}
\author{Mostafa El-Khamy$^{*}$ and  Robert J. McEliece$^{**}$ \\
Department of Electrical Engineering\\
California Institute of Technology, Pasadena CA 91125 USA\\
\begin{tabular}{c c}
$^{*}$E-mail: mostafa@systems.caltech.edu & $^{**}$E-mail: rjm@systems.caltech.edu \\
\end{tabular}}
\newtheorem{thm}{Theorem}
\newtheorem{cor}[thm]{Corollary}

\newtheorem{alg}{Algorithm}

\usepackage{amsmath}
\usepackage{amssymb}
\usepackage{graphicx}
\usepackage{array}
\begin{document}
\maketitle
\sloppy
\begin{abstract}
In this paper, we present an iterative soft-decision decoding
algorithm for Reed-Solomon codes offering both complexity and
performance advantages over previously known decoding algorithms.
Our algorithm is a list decoding algorithm which combines two
powerful soft decision decoding techniques which were previously
regarded in the literature as competitive, namely, the Koetter-Vardy
algebraic soft-decision decoding algorithm and belief-propagation
based on adaptive parity check matrices, recently proposed by Jiang
and Narayanan. Building on the Jiang-Narayanan algorithm, we present
a belief-propagation based algorithm with a significant reduction in
computational complexity. We introduce the concept of using a
belief-propagation based decoder to enhance the soft-input
information prior to decoding with an algebraic soft-decision
decoder. Our algorithm can also be viewed as an interpolation
multiplicity assignment scheme for algebraic soft-decision decoding
of Reed-Solomon codes.
\end{abstract}
\section{Introduction}
Reed-Solomon (RS) codes \cite{RS60} are among the most celebrated
forward error correcting codes. The RS codes are currently used in a
wide variety of applications, ranging from satellite communications
to data storage systems. Reed-Solomon codes have been adopted as
outer codes in the 3G wireless standard, CDMA2000 high-rate
broadcast packet data air interface \cite{ARBCDMA}, and are expected
to be used as outer codes in concatenated coding schemes for future
4G wireless systems.

Maximum likelihood (ML) decoding of linear codes, in general, and RS
codes, in particular, is NP-hard \cite{BMT78, GurVardy}. It remains
an open problem to find polynomial-time decoding algorithms with
near ML performance.
 A soft-decision ML decoding algorithm was proposed by Vardy and Be'ery
 \cite{VardyB91}. Further modifications of this algorithm were
 also studied \cite{PVucetic02}.
  Guruswami and Sudan (GS) \cite{GS} \cite{S97}
invented a polynomial-time list
 decoding algorithm for RS codes capable of correcting beyond half the
 minimum distance of the code.
 Koetter and Vardy (KV) \cite{KV03} developed an algebraic soft-decision decoding
 (ASD) algorithm for RS codes based on a multiplicity assignment scheme for the GS algorithm.
Alternative ASD algorithms, such as the Gaussian approximation
algorithm by Parvaresh and Vardy \cite{PV03} and the algorithm by
El-Khamy and McEliece based on the Chernoff
bound\cite{ElkMcDimacs,ElKHMc04}, have better performance.

   Jiang and Narayanan (JN) developed an iterative algorithm based on belief propagation
    for soft decoding of RS codes  \cite{JN04, JN04b}. This algorithm
   compares favorably with other soft decision
     decoding algorithms for RS codes and is a major step towards
    message passing decoding algorithms for  RS codes.
In the JN algorithm, belief propagation is run on an \emph{adapted}
parity check matrix where the columns in the parity-check matrix
corresponding to the least reliable independent bits are reduced to
an identity submatrix \cite{JN04, JN04b}. The order statistics
decoding algorithm by Fossorier and Lin \cite{FossLin95} also sorts
the received bits with respect to their reliabilities and reduces
the columns in the generator matrix corresponding to the most
reliable bits to an identity submatrix. This matrix is then used to
generate (permuted) codewords using the most reliable bits.
 Other soft-decoding algorithms
for RS codes include the generalized minimum distance (GMD) decoding
algorithm introduced by Forney \cite{For66}, the Chase II algorithm
\cite{Chase72}, the combined Chase II-GMD algorithm
\cite{TanFosLin01} and successive erasure-error decoding
\cite{HuLin03}.

   In this paper, we develop an algebraic soft-decision list
   decoding algorithm based on the idea that belief propagation-based algorithms could be deployed
   to improve the reliability of the symbols that is then utilized by an interpolation multiplicity
   assignment algorithm. Our algorithm combines the KV and the JN algorithms.
   An outline of the paper is as follows.
   Some preliminaries are given in section \ref{Pr}.
In section \ref{asd}, we briefly review algebraic soft-decoding
algorithms, in general, and the KV algorithm, in particular. The JN
algorithm is explained in the context of this paper in section
\ref{ABP}. Some modifications to the JN algorithm are introduced in section \ref{MJN}.
One of the main contributions in this paper, the
iterative algebraic soft-decision list decoding algorithm, is presented in
section \ref{ABPASDALG}. Another main contribution, a low complexity
   algorithm based on the JN algorithm, is presented in section \ref{LCABPASD}.
    Some discussions as well as some numerical results are presented in section \ref{num}.
   Finally, we conclude the paper
   in section \ref{conc} and suggest future research directions.

   \section{\label{Pr} Preliminaries}

   Throughout this paper, $\mbit{d}=[d_0,d_1,...,d_{k-1}]$ will denote a $k$ dimensional vector over $F_q$
   where
$F_q$ is the finite
 field of $q$ elements. $\mathcal{C}$ will denote an $(n,k)$ RS
 code. An $(n,k)$ RS codeword
  $\mbit{u}=[u_0,u_1,..,u_{n-1}]$ could be generated by
evaluating the data polynomial $D(x)=\sum_{i=0}^{k-1}d_i x^i$ at
 $n$ elements of the field
composing a set, called the support set of the code. This set is
vital for the operation of the Guruswami-Sudan algorithm. Let
$\alpha$ be a primitive element in $F_q$. Since
 the polynomial $U(x)=\sum_{i=0}^{n-1} u_i x^i $
associated with the codeword $\mbit{u} \in \mathcal{C}$ generated by polynomial
evaluation has $\alpha, \alpha^2,..,\alpha^{n-k}$ as zeros \cite{McSln},
a valid parity check matrix for  $\mathcal{C}$ is \cite{Mc02book}
\eq{\label{parity check} \mathcal{H}= \left[
\begin{array}{c c c c }
1&\alpha& \ldots&\alpha^{n-1} \\
1 &\alpha^2& \ldots&\alpha^{2(n-1)}\\
\vdots&\vdots&\ldots&\vdots\\
1 &\alpha^{n-k}& \ldots&\alpha^{(n-k)(n-1)}
\end{array}
 \right].
}

The redundancy of the code's binary
image will be denoted by $\tilde{r}$ where
$\tilde{r}=\tilde{n}-\tilde{k}$ and
 $\tilde{n}=mn$ and $\tilde{k}=mk$.
The results in this paper assume that the binary image and the
corresponding binary parity check matrix are of the form described
here. Let $p(x)$ be a primitive polynomial in $F_2[x]$ and $C$ be
its companion matrix \cite{Horn85}. The companion matrix is an $ m
\times m$ binary matrix. Since the mapping $\alpha^i \leftrightarrow
C^i$, $\{i=0,1,2,..\}$ induces a field isomorphism, an $\tilde{r}
\times \tilde{n}$ binary parity check matrix $H$ is obtained by
replacing every element $\alpha^i$ in the parity check matrix
$\mathcal{H}$ by its corresponding $m \times m$ matrix
$\mathcal{C}^i$. The binary image $\mbit{b}$, such that
$H\mbit{b}^T=0$, is obtained by representing each element $u_j \in
F_{2^m}$ with $u_j=u_{j,0} + u_{j,1} \alpha +
...+u_{j,m-1}\alpha^{m-1}$ where ${\;\;u_{j,i} \in F_2}$.

An $q \times n$ array of real numbers will be denoted by
$W=[W_i(\b)]$, where $i=0,1,...,n-1$ and $\b \in F_q$. If $\mbit{u}$
is transmitted and the corresponding channel output is $\mbit{y}$,
then we denote the a-posteriori probabilities $Pr\{u_i=\b | y_i\}$
by $\Pi_i(\b)$.

\section{\label{asd} Algebraic Soft Decoding}

An algebraic soft decoder makes use of the soft information
available from the channel. Given the a-posteriori probability
matrix $\Pi$, a multiplicity assignment algorithm generates an $q
\times n$ \textit{multiplicity} matrix, $ M=[M_i(\beta)]$, of
non-negative integers. The \emph{interpolation cost} of $M$ is
defined to be \footnote{To prevent notational ambiguity, $\|x\|_1$
will denote the magnitude of $x$.} $|M| \Def {1 \over 2}
\sum_{i=0}^{n-1} \sum_{\b \in F_q} W_i(\b)\left(W_i(\b) +1\right)$
 and the \emph{score} of $\mbit{u}$ with respect to $M$ is $\Inner(\mbit{u} ,M)\Def \sum_{i=0}^{n-1}
M_i(u_i).$
This multiplicity matrix is then passed to a (modified) GS
algorithm consisting of two main steps \cite{GS,Mc03tech}
\begin{enumerate}
    \item \emph{Interpolation:} Construct a bivariate polynomial, $Q(x,y)$, of
minimum $(1,k-1)$ weighted degree that passes through each of the
points $ (T_i, \beta) $ with multiplicity $M_i(\beta)$, where $\beta \in F_q$ and $i=0,1,..,n-1$.
    \item \emph{Factorization:} Find all linear factors
$(y-G(x))|Q(x,y)$ where $G(x)$ is a polynomial of degree less than
$k$. Each such polynomial $G(x)$ is placed on the list.
\end{enumerate}
A solution to the interpolation problem exists if $|M|$ is strictly
less than the number of monomials in $Q$ such that $Q$ is of minimal
$(1,k-1)$ weighted degree  $\Delta_{k-1}(|M|)$ \cite{NH00}. A
sufficient condition  for a codeword $\mbit{u}$ to be on the GS
generated list is \cite{GS,KV03},
\begin{equation} \label{cond}
\Inner(\mbit{u} ,M)> \Delta_{k-1}(|M|),
\end{equation} where $ \Delta_{v}(\gamma)=\left \lfloor \frac{\gamma}{m} +
\frac{v(m-1)}{2} \right \rfloor $ for $ m=\left\lfloor\sqrt{\frac{2
\gamma}{v}+\frac{1}{4} }+ \frac{1}{2} \right\rfloor$
\cite{ElkMcDimacs}. In case the cost tends to infinity, the
sufficient condition is \cite{KV03, ElkMcDimacs}
\eq{\label{suffM}\frac{\Inner(\mbit{u} ,M)}{\|M\|_2}> \sqrt{k-1}.}

In this paragraph, we briefly review well-known ASD algorithms. For
more details, we refer the readers to the given references.
 The KV algorithm maximizes the mean of the score.
  A reduced complexity KV algorithm constructs the multiplicity matrix $M$ as follows \cite{KV03,Gross03b}
\begin{equation} \label{kv1}
M_i(\b)=\lfloor \lambda \Pi_i(\b) \rfloor, \end{equation} where
$\lambda> 0$ is a complexity parameter determined by $|M|$. For
$|M|= \gamma$, it can be shown that
$\lambda=(-1+\sqrt{1+8\gamma/n})/2$. Other algorithms \cite{PV03}
and \cite{ElkMcDimacs} minimize the error probability directly. The
algorithm of \cite{PV03} (Gauss) assumes a Gaussian distribution of
the score,
 while that of \cite{ElkMcDimacs} (Chernoff)
minimizes a Chernoff bound on the error probability. The later appears to have the best performance.

\section{\label{ABP} Adaptive Belief Propagation}
Gallager devised an iterative algorithm for decoding his low-density
parity check (LDPC) codes \cite{Gal63}. This algorithm was the first
appearance in the literature of what we now call belief propagation
(BP). Recall that $H$ is the parity check matrix associated with the
binary image of the RS code. It has $\tilde{r}$ rows corresponding
to the check nodes and $\tilde{n}$ columns corresponding to the
variable nodes (transmitted bits). $H_{i,j}$ will denote the element
in the $i^{th}$ row and $j^{th}$ column of $H$. Define the sets,
$J(i)\Def \{j\;|\;H_{i,j}=1\}$ and $I(j) \Def \{i\;|\;H_{i,j}=1\}$.
Define $Q_{i,j}$ to be the log-likelihood ratio (LLR) of the $j$th
symbol, $u_j$,
 given the information about all
parity check nodes except node $i$ and $R_{i,j}$ to be the LLR
that check node $i$ is satisfied when $u_j$ is fixed to $0$ and
$1$ respectively. 
 Given the vector $\mbit{\Lambda^{in}}$ of initial LLRs,
 the BP algorithm outputs the extrinsic LLRs $\mbit{\Lambda^{x}}$ as described below
  \cite{McMackay98}\cite{HagOP96}.
 \begin{alg}{Damped Log Belief Propagation (LBP)}\\
{For all  $(i,j)$ such that $H_{i,j}=1$:\\
 {\textbf{Initialization:}
$Q_{i,j}=\Lambda^{in}_{j} $ \\
\textbf{DO}}}

{\textbf{Horizontal Step:} \eqar{\nonumber R_{i,j}&=&
\log\left(\frac{1+\prod_{k \in J(i)\setminus j}\tanh(Q_{i, k}/2)}
{1-\prod_{k \in J(i)\setminus j}\tanh(Q_{i, k}/2)}\right) \\
&=& \label{HS2} 2 \tanh^{-1}\left(\prod_{k \in J(i)\setminus j}\tanh(Q_{i, k}/2)\right)}}

{ \textbf{Vertical Step:}
\[Q_{i,j}=\Lambda^{in}_j+\theta \sum_{k \in I(j)\setminus i} R_{k, j}\]
\textbf{While} stopping criterion is not met.}

 {\textbf{Extrinsic Information:}
$\Lambda^{x}_j=\sum_{k \in I(j)} R_{k,j}$.}
\end{alg}
The factor $\theta$ is termed the \emph{vertical step damping factor} and
$ 0 < \theta \leq 1$. The magnitude of $\theta$ is determined
by our level of confidence about the extrinsic information. In our implementations,
$\theta$ is $0.5$. Eq. \ref{HS2} is specifically useful for fast hardware
implementations where the $\tanh$ function will be quantized to a reasonable accuracy and implemented as a
lookup table. In our implementation, damped LBP is run for a small
number of iterations on a fixed parity check matrix, so the stopping
criterion is the number of iterations. In case that only one LBP iteration is run on the parity check matrix,
the vertical step is eliminated.

Following we describe the JN algorithm \cite{JN04, JN04b}, which builds on the BP algorithm.
In the JN algorithm, BP is run on the parity check matrix after reducing its independent columns
corresponding
to the least reliable bits to an identity submatrix. We will refer to such a class of algorithms,
that adapt the
parity check matrix before running BP, by adaptive belief propagation (ABP).

\begin{alg}\label{AlgABP}{The JN Algorithm}\\
 {\textbf{
Initialization:} $\mbit{\Lambda^{p}}:=\mbit{\Lambda^{ch}}$\\
 \textbf{DO}}
\begin{enumerate}
    \item
{Sort $\mbit{\Lambda^{p}}$ in ascending order of magnitude and store
the sorting index. The resulting vector of sorted LLRs is
\[\mbit{\Lambda^{in}}=[\Lambda^{in}_{1},\Lambda^{in}_2,...,\Lambda_{nm}^{in}],\]
$\|\Lambda^{in}_k\|_1 \leq \|\Lambda^{in}_{k+1}\|_1$ for $k=1,2,...,nm-1$
and $\mbit{\Lambda^{in}}=P\mbit{\Lambda^{p}}$ where $P$ defines a permutation
matrix.}
    \item
 {Rearrange the columns of the binary parity check matrix $H$ to
form a new matrix $H_P$ where the rearrangement is defined by the
permutation $P$.}
    \item
{Perform Gaussian elimination (GE) on the matrix $H_P$ from left to
right. GE will reduce the first independent $(n-k)m$ columns in
$H_P$ to an identity sub-matrix. The columns which are dependent on
previously reduced columns will remain intact. Let this new matrix
be $\hat{H}_P$.}
    \item
{Run log BP on the parity check matrix $\hat{H}_P$ with initial LLRs
$\mbit{\Lambda^{in}}$ for a maximum number of iterations $It_H$ and
a vertical step damping factor  $\theta$. The log BP algorithm
outputs extrinsic LLRs $\mbit{\Lambda^{x}}.$}
    \item
{Update the LLRs, $\mbit{\Lambda^q}=\mbit{\Lambda^{in}}+\alpha_1
\mbit{\Lambda^{x}}$ and $\mbit{\Lambda^{p}}:=P^{-1}\mbit{\Lambda^q}$
where $0<\alpha_1\leq 1$ is called the ABP damping factor and
$P^{-1}$ is the inverse of $P$.}
\item \label{Decstep} {Decode using $\Lambda^p$ as an input to the decoding algorithm $D$.}
\end{enumerate}
{\textbf{While} Stopping criterion not satisfied.}\end{alg}

The JN algorithm assumed that the decoder $D$ is one of the following hard-decision decoders:
\begin{itemize}
\item HD: Perform hard-decisions on the updated LLRs, $\mbit{\hat{u}}
= (1-\mathrm{sign} (\mbit{\Lambda^p}))/2.$ If $H\mbit{\hat{u}}^T=0 $, then
a decoding success is signaled.
\item BM: Run a bounded minimum distance decoder such as the Berlekamp-Massey (BM) algorithm on the LLRs
after hard-decisions. If the BM algorithm finds a codeword, a decoding success is signaled.
\end{itemize}
The performance largely depends on the decoder $D$ 
and the stopping criterion used. This is discussed in the following section.

\section{Modifications to the JN Algorithm \label{MJN}}
The stopping criterion deployed in the JN algorithm is as follows
\cite{JN04b}:
\begin{itemize}
\item Stop if a decoding success is signaled by the decoder $D$ or if the number of iterations
is equal to the maximum number of iterations, $N_1$.
\end{itemize}

We propose a list-decoding stopping criterion in which a list of codewords is iteratively generated.
The list-decoding stopping criterion is as follows
\begin{itemize}
\item If a decoding success is signaled by the decoder $D$, add the decoded codeword
to a \emph{global} list of codewords. Stop if the number of iterations is equal to the maximum
number of iterations, $N_1$.
\end{itemize}
If more than one codeword is on the global list of codewords, then the list-decoder's output is the
codeword which is at the minimum Euclidean distance from the received vector.
Alternatively, one could only save the codeword with the largest conditional
probability, given the received vector. This codeword would be the candidate for the list
decoder's output
when the iteration loop terminates.

The advantage of our proposed list-decoding stopping criterion over
the stopping criterion in the JN algorithm is emphasized in the case
of higher rate codes, where the decoder error probability is
relatively high. Given a decoding algorithm $D$, the JN ABP
algorithm may result in updating the received vector to lie in the
decoding region of an erroneous codeword. However, running more
iterations of the JN ABP algorithm may move the updated received
vector into the decoding sphere of the transmitted codeword. The
decoding algorithm $D$ should also be run on the channel LLRs before
any ABP iteration is carried out. If the decoder succeeds to find a
codeword, it is added to the list.

Jiang and Narayanan \cite{JN04} proposed running $N_2$ parallel decoders (outer iterations),
 each with the JN stopping criterion and a maximum of $N_1$ inner iterations.
Each
one of these $N_2$ iterations (decoders) starts with a different random permutation
of the sorted channel LLRs in the first inner iteration. The outputs of these
$N_2$ decoders form a list of at most $N_2$ codewords. If each of these $N_2$ decoders
succeeds to find a codeword, the closest codeword to the received vector is chosen.
We also run $N_2$ parallel decoders (outer iterations), each with the list-decoding stopping criterion,
to form a global list of at most $N_1 N_2$
codewords. We propose doing the initial sorting of the channel LLRs in a systematic way
to ensure that most bits
will have a chance of being in the
identity sub-matrix of the adapted parity check matrix. The improved
performance achieved by these restarts could be explained by
reasoning that if a higher reliability bit is in error, then it has a higher
chance of being corrected if its corresponding column in the parity check
matrix is in the sparse identity submatrix.

Let $z=\lfloor
\tilde{n}/N_2\rfloor$, then at the $(j+1)^{th}$ outer iteration, $j>0$,
the initial LLR vector at the first inner iteration is 
\eq{[\Lambda^{in}_{jz+1},..,\Lambda^{in}_{(j+1)z},\Lambda^{in}_{1},...,
\Lambda^{in}_{jz},\Lambda^{in}_{(j+1)z+1},...,\Lambda^{in}_{\tilde{n}}],}
where $\mbit{\Lambda^{in}}$ is the vector of sorted channel LLRs.
The columns of $H_P$ will also be rearranged according to the same
permuatation. If $(j+1)z \leq \tilde{r}$, then it is less likely
that this initial permutation will introduce new columns into the
identity submatrix other than those which existed in the first outer
iteration. After the first outer iteration, it is thus recommended
to continue with the $(j+1)$th outer iteration such that $(j+1) >
\tilde{r}/z$.

Another modification that could improve the performance of the JN
algorithm is to run a small number of iterations of damped log
belief propagation on the same parity check matrix. Although belief
propagation is not exact due to the cycles in the associated Tanner
graph, running a very small number of iterations of belief
propagation is very effective \cite{Yed03}. Observing that the
complexity of belief propagation is much lower than that of Gaussian
elimination, one gets a performance enhancement at  a slightly
increased complexity.

Throughout the remaining of this paper,
we will refer to the modified JN algorithm with a list decoding
stopping criterion, as well as with the other modifications introduced in this section,
by ABP-BM if the decoding algorithm $D$ is BM. Similarly, if the decoding algorithm was HD,
the algorithm is referred to by ABP-HD. One of the main contribution in this
paper, the utilization of the a-posteriori probabilities at the output of
the ABP algorithm as the soft information input to an ASD
algorithm, is presented in the following section.

\section{The Hybrid ABP-ASD List Decoding Algorithm \label{ABPASDALG}}

Koetter and Vardy ~\cite{KV03} point out that it is hard to maximize
the mean of the \emph{score}
 with respect to the
to the true channel a-posteriori probabilities.
Previous multiplicity assignment algorithms ~\cite{KV03, PV03, ElkMcDimacs}
assumed approximate a-posteriori probabilities.
The problem is simplified by assuming that the transmitted
codeword is drawn uniformly from $F_q^n$. Also, the $n$ received symbols are assumed
to be independent and thus be assumed to be uniformly distributed. In such a case, the a-posteriori
probabilities are approximated to be a scaling of the channel transition probabilities,
\eq{\Pi^{ch}_i(\b)=\frac{Pr\{y_i|u_i=\b\}}{\sum_{\omega \in F_q} Pr\{y_i|u_i=\omega\}}.} However,
from the maximum distance separable (MDS) property of RS codes any $k$ symbols (only)
are $k$-wise independent
 and could be treated as information symbols and thus uniformly distributed.
Thus these assumptions are more valid for higher rate codes and for
memoryless channels. It is well known that belief propagation
algorithms improve the reliability of the symbols by taking into
account the geometry of the code and the correlation between symbols
(see for example ~\cite{McMackay98}.) Due to the dense nature of the
parity check matrix of the binary image of RS codes, running belief
propagation directly will not result in a good performance. Because
the Tanner graph associated with the parity check matrix of the
binary image of RS codes has cycles, the marginals passed by the
(log) belief propagation algorithm are no longer independent and the
information starts to propagate
in the loops. 

Jiang and Narayanan \cite{JN04b} proposed a solution to this problem
by adapting the parity check matrix after each iteration. When
updating the check node reliabilities $R_{i,j}$ (see (\ref{HS2}))
corresponding to a pivot in a single weight column, the information
$Q_{i,j}$ from any of the least reliable independent bits does not
enter into the summation. One reason for the success of ABP is that
the reliabilities of the least reliable bits are updated by only
passing the information from the more reliable bits to them. An
analytical model for belief propagation on adaptive parity check
matrices was recently proposed \cite{Ah04}.

 Our ABP-ASD algorithm is summarized by the following chain,
\eq{{\mbit{u}}  \to \Pi^{ch} \buildrel ABP \over \longrightarrow {\hat{\Pi}}
\underbrace{\buildrel \mathcal{A }\over \longrightarrow {M}  \to}_{ASD} {\mbit{\hat{u}}}
\label{eq:Markov1},}
where $\mbit{u}$ is the transmitted codeword, $\mathcal{A}$ is a multiplicity assignment algorithm,
 $M$ is the multiplicity matrix and $\mbit{\hat{u}}$ is the decoder output.
In particular, the ABP-ASD list decoder is implemented by deploying
 the list decoder stopping criterion, proposed in the previous section, with
an ASD decoding algorithm $D$ (see Alg. \ref{AlgABP}):
\begin{itemize}
\item ASD:  Using $\mbit{\Lambda^{p}}$ generate an $q \times
n$ reliability matrix $\hat{\Pi}$ which is then used as an input
to an multiplicity assignment algorithm to generate multiplicities
according to the required interpolation cost.
 This multiplicity matrix is passed to the (modified) GS list
decoding algorithm. If the generated codeword list is not
empty, the list of codewords is augmented to the \emph{global} list of codewords.
If only one codeword is required, the codeword with the
highest reliability with respect to the channel LLR's
$\mbit{\Lambda^{ch}}$ is added to the global list.

\end{itemize}
\begin{figure}
\centering
\includegraphics[width=3.5in, height=3.5in]{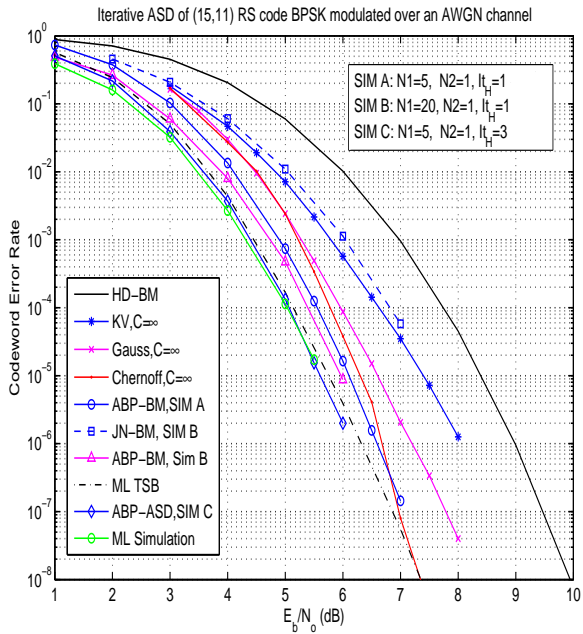}
\label{ASD1511c} \caption{The performance of iterative ASD of
(15,11) RS code, which is BPSK modulated and transmitted over an
AWGN channel, is compared to that of other ASD algorithms and ABP-BM
list decoding}
\end{figure}

In this paper, the KV algorithm is used as the multiplicity
assignment scheme. More efficient but more complex MA schemes could
also be used \cite{ElkMcDimacs}. The joint ABP-ASD algorithm
corrects decoder failures (the received word does not lie in the
decoding region centered around any codeword) of the ASD decoder
$D$, by iteratively enhancing the reliabilities of the received
word, and thus moving the received word into the decoding region
around a certain codeword. The decoding region in turn depends on
the algorithm $D$ and the designed interpolation cost. Furthermore,
it attempts to eliminate decoder errors (the decoded codeword is not
the transmitted codeword) by iteratively adding codewords to the
global list of codewords and choosing the most probable one.

Since ASD is inherently a list decoding algorithm with a larger
decoding region, it is expected that ABP-ASD  outperforms ABP-HD and
ABP-BM. Since our algorithm transforms the channel LLRs into
interpolation multiplicities for the GS algorithm, then, by
definition, it is an interpolation multiplicity assignment algorithm
for ASD.

The ABP-ASD algorithm has a polynomial-time complexity. The ABP step
involves $o(\tilde{n}^2)$ floating point operations, for sorting and
BP, and $o( min(\tilde{k}^2, \tilde{r}^2)\; \tilde{n})$ binary
operations for GE \cite{JN04}. As for ASD, the KV MA algorithm (see
(\ref{kv1})) has a time complexity of $O(n^2)$. An efficient
algorithm for solving the interpolation problem is Koetter's
algorithm \cite{Mc03tech} with a time complexity of $O(n^2
\lambda^4)$. A reduced complexity interpolation algorithm is given
in \cite{NH00}. Roth and Ruckenstein \cite{RR00} proposed an
efficient factorization algorithm with a time complexity $O((l
\log^2 l) k (n+l \log q))$, where $l$ is an upper bound on the ASD's
list size and is determined by $\lambda$.

\begin{figure}
\centering
\includegraphics[width=3.5in, height=3.5in]{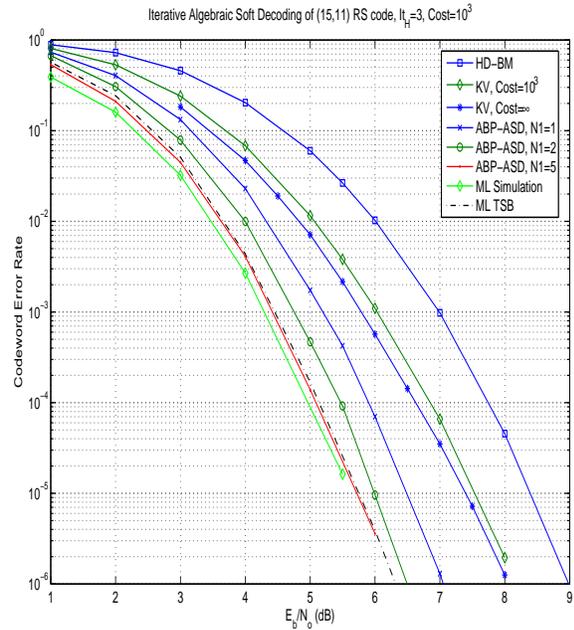}
\label{ASD1511finite} \caption{The performance of iterative ASD of
(15,11) RS code, which is BPSK modulated and transmitted over an
AWGN channel, is shown for a finite interpolation cost of $10^3$ and
different iteration numbers.}
\end{figure}

\section{A Low Complexity ABP Algorithm \label{LCABPASD}}
Most of the complexity of adaptive belief propagation lies in row reducing the
 binary parity check matrix
(after rearranging the columns according to the permutation $P$).
To reduce the complexity one could make use of the columns already reduced in the previous iteration.

We will use the same notation as in Alg. $2$ with a subscript $j$ to denote the
values at iteration $j$.
 For example, the vector of sorted LLRs at the $j$th iteration is $\mbit{\Lambda^{in}_j}$.
 Define $P_j(H)$ to be the matrix obtained when the columns of the parity check matrix $H$
  are permuted according to the permutation $P_j$ at the $j$th iteration.
 $GE(H)$ will be the reduced matrix (with an identity submatrix) after Gaussian elimination
  is carried out on the matrix $H$.

 Let $R_j \Def \{t: \; t$th column of $H$ was reduced to a column of unit weight in $GE(P_j(H))\}$.
 It is clear that the cardinality of $R_j$ is $\tilde{r}$.
Now assume that log BP is run and that the LLRs are updated and inverse permuted to get $\mbit{\Lambda^p_j}$
 (step 5 in Alg. 2).
The set of indices of the $\tilde{r}$ (independent) LLRs in $\mbit{\Lambda^p_j}$ with the smallest magnitude
will be denoted by  $S_{j+1}$.
By definition, $P_{j+1}$ is the permutation that sorts the LLRs in $\mbit{\Lambda^{p}_j}$
in ascending order according to their magnitude to get $\mbit{\Lambda^{in}_{j+1}}$.
The set $U_{j+1} \Def R_j\bigcap S_{j+1}$ is thus the set of indices of bits which are among
the least reliable independent bits at the $(j+1)$th iteration and whose corresponding columns in the reduced parity check matrix
 at the previous iteration were in the identity submatrix.

The algorithm is modified such that GE will be run on the matrix
whose left most columns are those corresponding to $U_{j+1}$. To
construct the identity submatrix, these columns may only require row
permutations for arranging the pivots (ones) on the diagonal. Note
that these permutations may have also been required when running GE
on $P_{j+1}(H)$. Only a small fraction of the columns will need to
be reduced to unit weight leading to a large reduction in the GE
computational complexity. Also note that what matters is that a
column corresponding to a bit with low reliability lies in the
identity (sparse) submatrix and not its position within the
submatrix. This is justified by the fact that the update rules for
all the LLRs corresponding to columns in the identity submatrix are
the same. Thus provided that the first $\tilde{r}$ columns in
$P_{j+1}(H)$ are independent, changing their order does
 not alter the performance of the ABP
algorithm. To summarize the proposed reduced complexity ABP algorithm can be stated as follows:

\begin{figure}
\centering
\includegraphics[width=3in, height=3.5in]{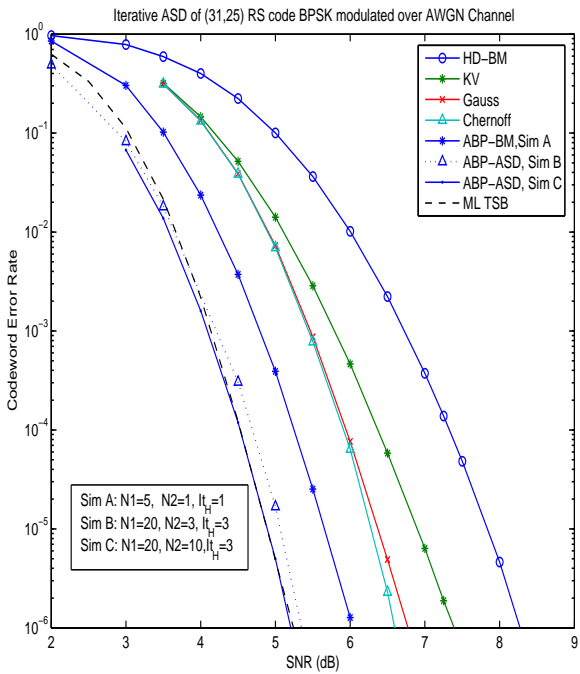}
\label{ASD3125} \caption{ABP-ASD list decoding of the (31,25) RS
code transmitted over an AWGN with BPSK modulation.}
\end{figure}

\begin{alg}{Low Complexity Adaptive Belief Propagation}\\
 {\textbf{
Initialization:} $\mbit{\Lambda^{p}}:=\mbit{\Lambda^{ch}}, j=1$\\
 \textbf{DO\\}
{If $j=1$\\} {Proceed as in the first iteration of Alg. 2;
$\mbit{\Lambda^{in}_1}=\mbit{\Lambda^{in}}|_{Alg. 2}$, $P_1=P|_{Alg.
2}$,
 $\hat{H}_{1}=\hat{H}_{P}|_{Alg. 2}$ and
$\mbit{\Lambda^{q}_1}=\mbit{\Lambda^{q}}|_{Alg. 2}$.\\}
 {If $j>1$}}
\begin{enumerate}
    \item
{Sort the updated LLR vector  $\mbit{\Lambda^{q}_{j-1}}$ in
ascending order of the magnitude of its elements.
 Let $W'_j$ be the associated sorting
permutation matrix.}
  \item
 {Rearrange the columns of the binary parity check matrix $\hat{H}_{j-1}$ to
form a new matrix \eqn{Q'_j=W'_j(\hat{H}_{j-1}).}}
\item
 {\label{sortABP3}Rearrange the  most left $\tilde{r}$ columns of the binary parity check matrix
 $Q'_j$ such that the columns of unit weight are the most left columns. Let $W''_j$
 be the corresponding permutation matrix. (This could be done
 by sorting the first $\tilde{r}$ columns of $Q'_j$ in ascending
 order according to their weight.)
 Let the resulting matrix be \eqn{Q''_j=W''_j(Q'_j).}}
 \item {Permute the LLR vector; \eqn{\mbit{\Lambda^{in}_j}=
 P'_j\mbit{\Lambda^{q}_{j-1}},}
 where  $P'_j=W'_jW''_j$.}
    \item  {Update the (global) permutation matrix; \eqn{P_j=P'_j P_{j-1}.}}
    \item
{Run Gaussian elimination on the matrix $Q''_j$ from left to right;
\eqn{\hat{H}_{j}=GE(Q''_j).}}
    \item
{Run damped LBP on $\hat{H}_{j}$ with initial LLRs
$\mbit{\Lambda^{in}_j}$ for $It_H$ iterations. The output vector of
extrinsic LLRs is $\mbit{\Lambda^{x}_j}.$}
    \item
{Update the LLRs;
\eqn{\mbit{\Lambda^q_j}=\mbit{\Lambda_j^{in}}+\alpha_1
\mbit{\Lambda^{x}_j} \mbox{ and }
 \mbit{\Lambda^{p}_j}=P^{-1}_j\mbit{\Lambda^q_j}.}}
 \item  {Decode using $\mbit{\Lambda^p_j}$ as an the input to the decoding algorithm $D$.}
 \item {Increment $j$.}
\end{enumerate}
{\textbf{While} Stopping criterion not satisfied.}\end{alg}

\begin{figure}
\centering
\includegraphics[width=3.5in, height=3.5in]{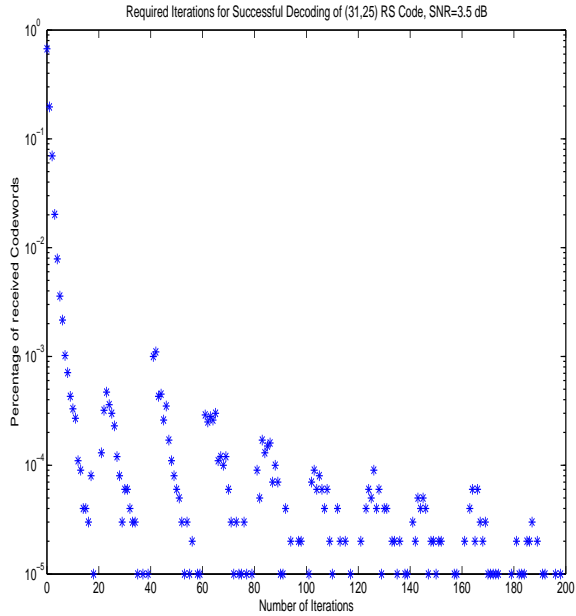}
\label{3125Hist} \caption{This histogram shows the percentage of
transmitted codewords successfully decoded versus the iteration
number at which the transmitted codeword was first successfully
added to the ABP-ASD list with $N1=20$ and $N2=10$. The $(31,25)$ RS
code is transmitted over an AWGN channel at an SNR of $3.5$ dB.}
\end{figure}

The algorithm as described above iteratively updates a global permutation matrix and avoids
inverse permuting the row-reduced parity check matrix in each iteration. The implementation
of the algorithm also assumes for simplicity that the columns in the parity check matrix corresponding to
the $\tilde{r}$ least reliable bits
are independent and could therefore be reduced to unit weight columns.
 It is also noticed that in practice the cardinality of $U_{j+1}$
is close to $\tilde{r}$ which means that the GE elimination complexity will be significant only in the first  iteration.

We will assume the favorable condition in which the most left
$\tilde{r}$ columns of an parity check matrix are independent. Taking into
account that the parity check matrix is a binary matrix, the maximum
number of binary operations required to reduce the first
$\tilde{r}$ columns to an identity submatrix in the JN algorithm (Alg. 2)  can be shown to be
\eq{\Theta_{GE}=
2\sum_{\alpha=1}^{\tilde{r}}(\tilde{r}-\alpha)(\tilde{n}-\alpha+1)<
\tilde{r}^2\tilde{n}- \tilde{r}\tilde{k}.} (It is assumed that the
two GE steps, elimination and back substitution, are symmetric).
Row permutation operations were neglected. Now assume that the cardinality of
$U_{j+1}$ is $\delta \tilde{r}$, where  $\delta\leq 1$.
 For the modified algorithm, only row permutations may be
required for the first $\delta \tilde{r}$ columns to arrange the
pivots on the diagonal of the identity submatrix. These permutations
may also be required for the JN algorithm. Then the relative
reduction in complexity is \eqar{\nonumber \frac{\mbox{$\Theta_{GE}$
in Alg. 2} - \mbox{$\Theta_{GE}$ in Alg. 3}}
{\mbox{$\Theta_{GE}$ in Alg. 2}}=\\
\nonumber \frac{\sum_{\alpha=1}^{\delta\tilde{r}}(\tilde{r}-\alpha)
(\tilde{n}-\alpha+1)}{\sum_{\alpha=1}^{\tilde{r}}(\tilde{r}-\alpha)(\tilde{n}-\alpha+1)}\approx
\\ \label{RR}
\frac{(\tilde{r}^2 \tilde{n})(2 \delta-\delta^2)-\delta
\tilde{r}\tilde{k}} {\tilde{r}^2 \tilde{n}-\tilde{r}\tilde{k}}
\approx 2 \delta-\delta^2.}

For example, if we assume that on average $\delta=0.5$,
 a simple calculation for the $(255,239)$ code over $F_{256}$ shows that the relative reduction
  in the complexity of
 the GE step is
 about $75\%$. In practice $\delta$ is close to one.
 Note that Alg. 3 does require sorting $\tilde{r}$ columns
of $Q'_j$ (see step (\ref{sortABP3})) according to their weight but the complexity is relatively small.

\section{Numerical Results and Discussions \label{num}}
In the next subsection, a fast simulation setup is described for ABP list decoding.
Bounds on the error probability of the ML decoder are then discussed.
We then show simulation results
for our algorithm.

\subsection{Fast Simulation Setup}
We describe a fast simulation setup for ABP with a list decoding stopping criterion.
One could avoid running the actual decoder
$D$ at each iteration and instead check whether the
transmitted codeword is on the list generated by the decoder
$D$. The stopping criterion would be modified such that the iterative decoding stops if
the transmitted codeword is on the list or if the maximum number of iterations is reached.
A decoding success is signaled if the transmitted codeword is on the list.

It is easy to see that this simulation setup is equivalent to
running
 the actual ABP list decoder for the maximum
number of iterations. Suppose that the received sequence results in
an maximum likelihood (ML) error, then it is very unlikely that the
decoder $D$ will correctly decode the received word at any
iteration. In case of an ML decoder success and the transmitted
codeword is added to the global list
 at a certain iteration,
 which presumably could be checked, then it would be the closest codeword to
 the received word and thus the list decoder's choice.
Thus for a fast implementation, a decoding success is signaled and iteration stops once
the transmitted codeword appears on the global
list.

In case that $D$ is a bounded minimum distance decoder such as the Berlekamp-Massey (BM) algorithm,
 the transmitted codeword would be on the global list if it is at a Hamming distance of $\leq \lfloor \frac{n-k}{2} \rfloor$
from the hard-decisioned (modified) LLRs. If $D$ is an ASD algorithm
that assigns the multiplicity matrix $M$, the transmitted codeword
is on the ASD's list (and thus the global list) if it satisfies the
sufficient conditions of (\ref{cond}) and (\ref{suffM}) for finite
and infinite interpolation costs respectively. It was shown in
\cite{KV03}, that simulating the KV algorithm by checking the
sufficient condition of (\ref{cond}) results in accurate results.
This is partially justified by the fact that on average, the ASD's
list size is one~\cite{Mc03list}.
 This is also justified by observing that if the ASD's list
is empty (a decoding failure), the condition (\ref{cond}) will not be satisfied.
However, if the list is nonempty
but the transmitted codeword is not on the list (a decoding error),
the condition will still not be satisfied for the transmitted codeword and a decoding error/failure is signaled.
However if the condition is satisfied, then this implies that the transmitted codeword is on the
ASD's list and thus a decoding success.

\subsection{Bounds on the ML error probability}
As important as it is to compare our algorithms with other algorithms, it is even more important
to compare it with the ultimate performance limits, which is that of the soft decision ML decoder.
 When transmitting
the binary image of RS codes over a channel, the performance of the
maximum likelihood decoder depends on the weight enumerator of the
transmitted binary image. The binary image of RS codes is not
unique, but depends on the
 basis used to represent the symbols as bits.
An average binary weight enumerator of RS codes could be derived by
assuming a binomial distribution of the bits in a non-zero symbol
\cite{ELKMcAller}. Based on the Poltyrev tangential sphere bound
(TSB)
 \cite{Polt94}
and the average binary weight enumerator, average bounds on the ML
error probability of RS codes over additive white Gaussian noise
(AWGN) channels were developed in \cite{ELKMcAller}
 and were shown to be tight.
We will refer to
this bound by ML-TSB. Alternatively the averaged
binary weight enumerator could be used in conjunction with other tight bounds such as the Divsalar simple bound
\cite{Divs}
to bound the ML error probability.

\subsection{Numerical Results}
In this subsection, we give some simulation results for our algorithm. As noted before, the
multiplicity
assignment algorithm used for ABP-ASD in the these simulations is the KV algorithm. $N2$ denotes
the number of outer iterations (parallel decoders) and $N1$ is the number of inner iterations in
each of these outer iterations.
\subsubsection{$(15,11)$ RS code over an AWGN channel}
A standard binary input AWGN channel is assumed where
the transmitted codewords are BPSK modulated.
In Fig. 1, we compare the performance of different decoding algorithms. 
HD-BM refers to the performance of a hard decision bounded minimum distance decoder such as the
BM algorithm.
The ABP-BM list decoding algorithm with $N1=5$ iterations and one iteration
of LBP on each parity check
matrix, $It_H=1$ (see step 4 in Alg. \ref{AlgABP}) has a coding gain of about
$2.5$ dB over HD-BM at a codeword error rate (CER) of $10^{-6}$.
Increasing the number of iterations to $N1=20$ iterations, we get
a slightly better performance. JN-BM refers to the JN algorithm with the JN stopping criterion and
a BM decoder. Due to the high decoder error probability of the $(15,11)$ code, ABP-BM, with the list decoder
stopping criterion, yields a much better performance than JN-BM.
The ABP-ASD list decoding algorithm outperforms all the previous algorithms with
only $5$ ABP iterations and with $It_H=3$. Comparing its performance with soft decision
ML decoding of the RS code, we see that ABP-ASD has a near ML performance with a performance gain of
about $3$ dB over HD-BM at a CER of $10^{-6}$.
(ML decoding was carried out by running the BCJR
algorithm on the trellis associated with the binary parity check matrix of the RS code \cite{MKan}.)
Moreover, the averaged TSB on the ML codeword error probability is shown to confirm that it is
a tight upper bound and that the ABP-ASD algorithm is near optimal for this code.

The performance of different ASD algorithms are compared for infinite interpolation
costs, the KV algorithm \cite{KV03}, the Gaussian approximation (Gauss) \cite{PV03}
and the Chernoff bound
algorithm (Chernoff) \cite{ElkMcDimacs}.
It is noted that the Chernoff bound algorithm has the best performance,
especially at the tail of error probability.
It is also interesting to compare the performance of ABP-ASD with other ASD MA algorithms.
It has about $2$ dB coding gain over the KV algorithm at
a CER of $10^{-6}$. As expected, the Chernoff
method has a comparable performance at the tail of the error
probability.

\begin{figure}
\centering
\includegraphics[width=3.5in, height=3.5in]{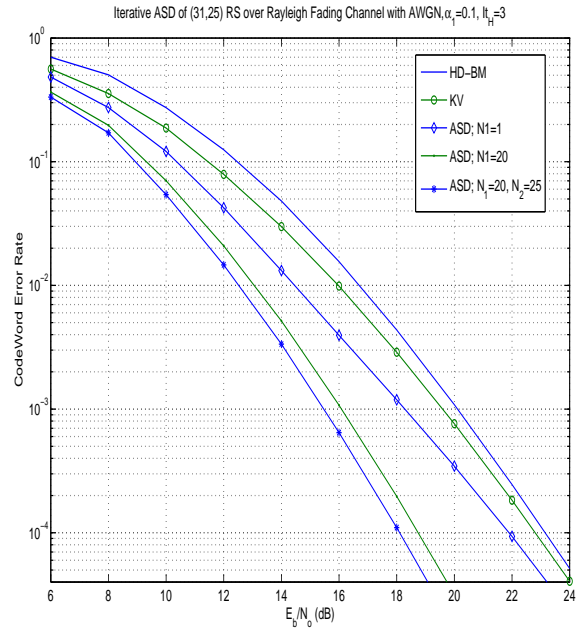}
\label{ASD3125Ray} \caption{The performance of the ABP-ASD decoding
of the $(31,25)$
 RS code over a Rayleigh fading channel with AWGN when the
channel information is unknown at the decoder.}
\end{figure}

The ABP algorithm used in the simulations shown in Fig. 1 is Alg. 2.
The performance of Alg. 3 was identical to that of Alg. 2. However,
the complexity is much less. The average $\delta$ (see (\ref{RR}))
averaged over all iterations was calculated versus the SNR. It was
observed that the ratio of the number of columns to be reduced in
Alg. 3 to that in Alg. 2 is about $0.1$
 ($\delta= 0.9$). This gives about a $99\%$ reduction in the Gaussian elimination complexity.
Thus only the first iteration or restart
suffers from an Gaussian elimination complexity if Alg. 3 is used.

Near ML decoding for the same code is also achieved by the ABP-ASD algorithm
with a finite cost of $10^3$ as shown in Fig. 2. 
Comparisons are
made between the possible
coding gains if the number of iterations is limited to $N1=1,2,5$.
With $5$ iterations, the performance gain over the KV algorithm, with the same interpolation cost, is nearly $1.8$ dB
at a CER of $10^{-5}$.
Comparing the ABP-ASD performance to that of Fig. 1, 
with infinite interpolation costs, we observe that
a small loss in performance results with reasonable finite interpolation costs. Unless otherwise stated, the remaining
simulations in this paper will assume infinite interpolation costs to show the potential of our algorithm.

 It is to be noted that in simulating the ABP-BM list decoder,
 the simulations using a real BM decoder
were identical to the simulations using the fast simulation setup described in this section.
To save simulation time, the curves shown here for ABP-ASD are generated using the fast simulation
setup. As is the case for ABP-BM, running the real ABP-ASD decoder will yield the same results.

\subsubsection{$(31,25)$ RS code over AWGN channel}
The arguments for the (15,11) RS code also carry over for the (31,25) RS code when BPSK modulated and
transmitted over an AWGN channel, as shown in Fig. 3. 
With only $5$ iterations, the ABP-BM list decoding algorithm
outperforms previous ASD algorithms.
The performance of ABP-ASD with $20$ inner iterations (N1)
 and 10 outer iterations (N2) is better than
the ML upper bound and has more than $3$ dB coding gain over the BM algorithm at an CER of $10^{-4}$. A favorable
performance is also obtained by only $3$ restarts (outer iterations). By comparing with Fig. 2 of \cite{TanFosLin01},
our ABP-ASD algorithm has about $1.6$ dB gain over the combined Chase II-GMD algorithm at an CER of $10^{-4}$.

To show the effectiveness of the restarts or outer iterations, we kept track of the iteration number at which
the ABP-ASD list decoder was first capable to successfully decode the received word. In other words, this is
the iteration when the transmitted codeword was first added to the ABP-ASD list. The percentage of
transmitted codewords which were first successfully decoded at a certain iteration is plotted versus the iteration
number in the histogram of Fig. 4. 
This is shown at a signal to noise ratio (SNR) of 3.5 dB and for $N1=20$ $N2=10$ with a total of $200$ iterations.
 At the beginning of each restart (every $20$ iterations)
there is a boost in the number of
codewords successfully decoded and this number declines again with increasing iterations.
The zeroth iteration corresponds to the KV algorithm. This histogram is also invaluable
for decoder design and could aid one to determine the designed number of iterations for a required CER.

\subsubsection{$(31,25)$ RS code over a Rayleigh Fading Channel}
As expected from the discussion in Sec. \ref{ABPASDALG}, the coding gain of ABP-ASD
is much more if the underlying channel model is
not memoryless. This is demonstrated in Fig. 5, 
 where an $(31,25)$ code is BPSK modulated over a
relatively fast Rayleigh fading channel with AWGN. 
 The Doppler frequency is equal to $50$ Hz and the
codeword duration is $0.02$ seconds. The coding gain of ABP-ASD over the KV algorithm at an CER of
$10^{-4}$ is nearly $5$ dB when the channel is unknown to both decoders.

\begin{figure}
\centering
\includegraphics[width=3.5in, height=3.5in]{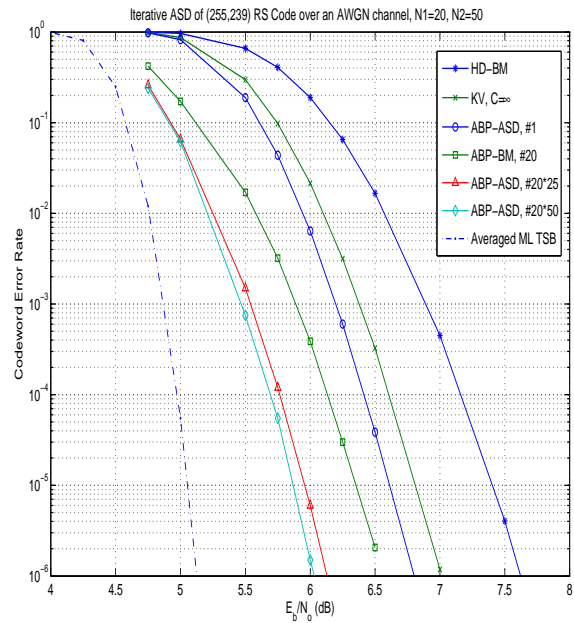}
\label{ASD255239} \caption{ The performance of the ABP-ASD decoding
of the $(255,239)$
 RS code over an AWGN channel with BPSK modulation.}
\end{figure}

\subsubsection{$(255,239)$ RS code over AWGN channel}
The performance of the ABP-ASD algorithm is also investigated for relatively long codes.
The $(255,239)$ code and its shortened version, the $(204,188)$ code,  are standards in many communication
systems.
The performance of the $(255,239)$ code over an AWGN channel is shown
in Fig. 6. 
By $20$ iterations of ABP-BM, one could achieve a coding gain of about $0.5$ dB over the KV algorithm.
 At an CER of $10^{-6}$,
after a total of  $25$ outer iterations (restarts), the coding gain of ABP-ASD over BM is about $1.5$ dB.
 An extra $0.1$ dB of coding gain is obtained with $25$ more outer iterations.
 Moreover, the performance of
 the ABP-ASD decoder is within $1$ dB of the averaged ML TSB.
\subsubsection{$(31,15)$ RS code over AWGN channel}
The performance of our algorithm is studied for the $(31,15)$ RS
code over an AWGN channel. The rate of this code is $0.48$. Because
this code is of relatively low rate, the HD-GS algorithm does
improve over the HD-BM bounded minimum distance decoding algorithm.
As seen from Fig. 7, ML soft-decision decoding offers about $4$ dB
coding gain over the hard decision GS algorithm and about $2.8$ dB
coding gain over the soft decision KV ASD algorithm at an CER of
$10^{-5}$. With $20$ iterations, ABP-BM list decoding improves over
the KV algorithm. As expected, ABP-ASD has a better performance for
the same number of iterations. With $10$ restarts, ABP-ASD has a
reasonable performance with about a $3$ dB coding gain over the BM
algorithm. Another $0.5$ dB of coding gain could be achieved by
increasing the number of iterations.

\subsubsection{General Observations} It is noticed that the coding gain between iterations decreases with the number of iterations.
It is also to be noted that the ABP-ASD list decoder  requires
running the KV ASD algorithm in each iteration. Running a number of
`plain-vanilla' ABP iterations without the ASD decoder and then
decoding using the ASD decoder (to reduce the complexity) will yield
a worse performance for the same number of iterations. The same
arguments also hold for the ABP-BM list decoding. A reasonable
performance is achieved by ABP-BM list decoding.
 By deploying the KV ASD algorithm, ABP-ASD list decoding
has significant coding gains over the KV ASD algorithm and other well known soft-decision decoding
 algorithms.
\begin{figure}
\centering
\includegraphics[width=3.5in, height=3.5in]{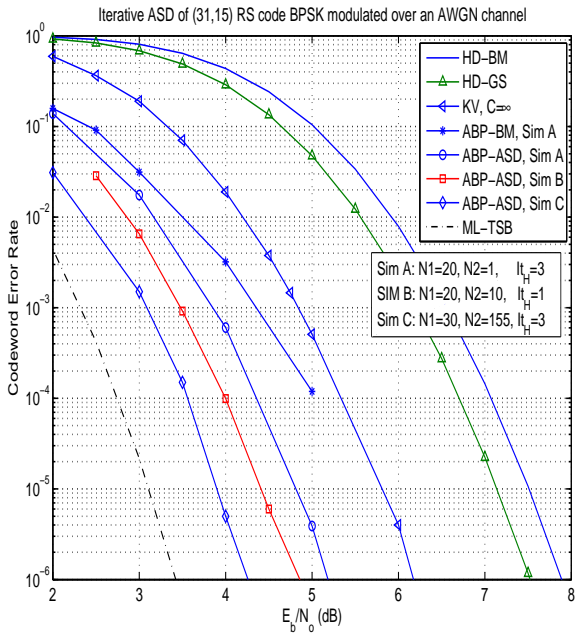}
\label{ASD3125} \caption{ABP-ASD list decoding of the (31,15) RS
code, of rate $0.48$, transmitted over an AWGN with BPSK
modulation.}
\end{figure}

\section{Conclusions \label{conc}}
In this paper, we proposed a list decoding algorithm for
soft-decision decoding of Reed-Solomon codes. Our algorithm is based
on enhancing the soft reliability channel information before passing
them to an algebraic soft-decision decoding algorithm. This was
achieved by deploying the Jiang and Narayanan algorithm, which runs
belief-propagation on an adapted parity check matrix. Using the
Koetter-Vardy algorithm as the algebraic soft-decision decoding
algorithm, our algorithm has impressive coding gains over previously
known soft-decision decoding algorithms for RS codes. By comparing
with averaged bounds on the performance of maximum likelihood
decoding of RS codes, we observe that our algorithm achieves a near
optimal performance for relatively short, high-rate codes. We
introduced some modifications over the JN algorithm that resulted in
better coding gains.
 We presented a low complexity adaptive belief-propagation algorithm, which results in
 a significant reduction in the computational complexity.
 The performance of our algorithm was studied for the cases when
 the interpolation cost of the algebraic soft-decision decoding algorithm is both finite and
 infinite.
 A small loss in coding gain results when using manageable interpolation costs.
 The coding gain of the presented algorithm is larger for channels with memory.
Our proposed algorithm could also be viewed as an interpolation
multiplicity assignment algorithm for algebraic-soft decoding.

The question remains whether the JN algorithm is the optimum way to
process the channel reliabilities before algebraic soft-decision
decoding. The KV algorithm was our ASD decoder of choice due to its
low complexity. Further investigations would be required to
determine the best ASD algorithm or, in general, soft-decision
decoding algorithm for joint belief-propagation list-decoding with
an eye on both the performance and computational complexity.

\section*{\hfil Acknowledgments}
The authors would like to thank J. Jiang and K. Narayanan for
providing an extended version of their paper \cite{JN04}. M.
El-Khamy is grateful to M. Kan for confirming many of the simulation
results in this paper. The authors gratefully acknowledge the
comments of the anonymous reviewers that have improved the
presentation of this paper.
\bibliographystyle{IEEEtran}
\bibliography{ArxivEkMef2.bbl}%








 \end{document}